\documentclass{ws-procs9x6}
\usepackage{amsmath,natbib}
\usepackage{theorem}

\newtheorem{mytheorem}{Theorem}
\newtheorem{mycase}{Case}
\newtheorem{mytask}{Task}
\newtheorem{myconjecture}{Conjecture}
     
\begin{document}

\title{The Chaotic Chameleon}

\author{Richard D. Gill}

\address{Mathematical Institute, University of Utrecht, Netherlands, \&\\
EURANDOM, Eindhoven, Netherlands\\
gill@math.uu.nl\\
http://www.math.uu.nl/people/gill
}

\maketitle

\abstracts{Various local hidden 
variables models for the singlet correlations
exploit the detection loophole, or other loopholes connected with
post-selection on coincident arrival times. 
I consider the connection with a probabilistic simulation
technique called rejection-sampling, and pose some natural questions
concerning what can be achieved and what cannot be achieved with
local (or distributed) rejection sampling. In particular a new and more
serious loophole, which we call the coincidence loophole, 
is introduced.}

\section{Introduction}

It has been well known since \citet{pearle}
that local realistic models can explain the singlet correlations
when these are determined on the basis of post-selected coincidences
rather than on pre-selected event pairs. These models are usually
felt to be unphysical and conspiratorial, and especially that they simply
exploit defects of present day detection apparatus (hence the name ``the detection loophole'').
However, Accardi, Ima\-fuku and Regoli 
(2002, 2003) (``\emph{the chameleon effect}''),
\citet{cath} (``\emph{the the chaotic ball effect}''), 
and others have argued that their models could make physical sense.
Further examples are provided by \citet{hp-a, hp-b, hp-c}, \citet{krack}, \citet{bs},
in many cases unwittingly. Already, \citet{gisin2} show that these models
can be simple and elegant, and should not be thought of as being artificial.

Accardi {\it et al.} (2002, 2003) furthermore insist that their work,
based on the \emph{chameleon effect},
has nothing to do with the so-called detection loophole. Rather, they claim
that the chameleon model is built on a fundamental legacy of measurement
of quantum systems, that there is also indeterminacy in whether or not
a particle gets measured at all, and when it gets measured. 
Furthermore, they focus entirely on 
perceived defects of the landmark paper Bell (1964),
where the incompatibility of the singlet correlations with
local realism was first established. Now Bell himself
became well aware of imperfections in his original work and 
in Bell (1981)  (reprinted in Bell, 1987), taking account of 
one and a half decades of  intense debate,  he explicitly 
elaborated on the experimental protocol which is 
necessary, before one can conclude from an experimental violation
of the Bell-CHSH inequality, that a local realistic explanation of the
observed phenomena is impossible. That protocol is not adhered
to by Accardi {\it et al.} (2002, 2003), nor (of course) by
any of the previously cited works in which local realistic violations of
Bell-CHSH inequalities are obtained.

It is a mathematical fact that ``chameleon model'' of the
type proposed by Accardi {\it et al.} (2002, 2003)  can be
converted into a ``detection loophole model'', and {\it vice-versa}.
This result has been independently obtained by
Takayuki Miyadera and Masanori Ohya, and by the present author
(unpublished).

In this paper I do not want to continue the philosophical debate, nor address 
questions of physical legitimacy of these models. 
See Gill (2003) for an overview, and in particular, for a discussion of the
option that \emph{quantum mechanics itself could prevent a succesfull
loophole-free experiment by preventing us from achieving the required
initial conditions}. Instead I would like to
extract a mathematical kernel from this literature, exposing some natural
open problems concerning properties of these models.
Possibly some answers are already known to experts on Bell-type
experiments and on distributed quantum computation.
I would especially like to pose these problems to experts in probability theory,
since the basic
renormalization  involved both in the chameleon model (under the name
of a ``form factor'') and in detection-loophole models, is well known in
probability theory under the name of \emph{rejection-sampling}.
From now I will use the language of Applied Probability: simulation, rejection-sampling,
and so on; and avoid reference to physics or philosophy.

The main new contribution of this paper is the discovery of a new loophole, which
we call the \emph{coincidence loophole}, which occurs when particle pairs are
selected on the basis of nearly coincident arrival times. It has recently been 
shown by \cite{lg} that this loophole is in a certain sense \emph{twice as serious}
as the well-known detection loophole.

\section{The Problem}

Suppose we want to simulate two random variables $X,Y$ from a joint probability
distribution depending on two parameters $a,b$. To fix ideas, let me give two key
examples:
\begin{mycase}\textbf{The Singlet Correlations.}
$X,Y$ are binary, taking the values\/ $\pm1$. The parameters\/ $a,b$ are two directions
in real, three dimensional space. We will represent them with two 
unit vectors in\/ $\mathbb R^3$ (two points on the unit sphere\/ $\mathbb S^2$). The joint density
of\/ $X,Y$ (their joint probability mass function) is
\begin{equation}
\Pr\nolimits_{a,b}\{X=x,Y=y\}~=~p(x,y;a,b)~=~\frac 14 \Bigl(1-xy\, a\cdot b\Bigr),
\end{equation}
where\/ $a\cdot b$ stands for the inner product of the unit vectors\/ $a$ and\/ $b$
and\/ $x,y=\pm 1$.
Note that the marginal laws of\/ $X$ and\/ $Y$ are both Bernoulli\/ {\rm ($\frac 12$)} on\/ $\{-1,+1\}$,
and their covariance equals their correlation equals\/ $-a\cdot b$.
In particular, the marginal law of\/ $X$ does not depend on $b$\/ nor that of $Y$\/
on\/ $a$.
\end{mycase}
\begin{mycase}\textbf{The Singlet Correlations Restricted.}
This is identical to the previous example except that we are only interested in\/ $a$
and\/ $b$ taking values in two particular, possibly different, finite sets of points on\/ $\mathbb S^2$.
\end{mycase}
\noindent Next I describe two different protocols for ``distributed Monte-Carlo simulation experiments'';
the difference is that one allows rejection sampling, the other does not.
The idea is that the random variables $X$ and $Y$ are going to be generated on
two different computers, and the inputs $a$, $b$ are only given to each computer separately.
The two computers are to generate dependent random variables, so they will start with having
some shared randomness between them. The programmer is allowed to start with any number
of random variables, distributed just how he likes, for this purpose. Cognoscenti will
realize that it suffices to have just one  random variable, uniformly distributed on the interval
$[0,1]$, or equivalently, an infinite sequence of fair independent coin tosses. There is no need
for the two computers to have access to further randomness---they may as well share 
everything they might ever need, separately or together, from the start.

The difference between the two protocols, or two tasks, 
is that the first has to get it right first time, or if
you prefer, with probability one. The second protocol is allowed to make mistakes, as
long as the mistakes are also ``distributed''. Another way to say this, is that we allow
``distributed rejection sampling''. Moreover, we allow the second protocol not to
be completely accurate. It might be, that the second protocol can be made more and
more accurate at the expense of a smaller and smaller acceptance (success) probability.
This is precisely what we want to study. Success probability and accuracy can both
depend on the parameters $a$ and $b$ so one will probably demand \emph{uniformly}
high success probability, and \emph{uniformly} good accuracy.
\begin{mytask}\textbf{Perfect Distributed Monte-Carlo.}
Construct a probability distribution of a random variable\/ $Z$, and two transformations\/
$f$ and\/ $g$ of\/ $Z$, each depending on one of the two parameters\/ $a$ and\/ $b$, such
that 
\begin{equation}
f(Z;a),g(Z;b)~~\sim~~X,Y\quad\text{ for all $a$, $b$}.
\end{equation}
\end{mytask}
\noindent The symbol `$\sim$' means `is jointly distributed as', and  $X,Y$ on the right hand side
come from the prespecified (or target) joint law with the given values of the 
parameters $a$ and $b$.

\begin{mytask}\textbf{Imperfect Distributed Rejection Sampling.}
As before, but there are two further transformations, let me call them\/ 
$D=\delta(Z,a)$ and\/ $E=\epsilon(Z;b)$, such that\/ $\delta$ and\/ $\epsilon$ take values\/
$1$ and\/ $0$ or if you like, {\rm \texttt{ACCEPT}} and  {\rm \texttt{REJECT}}, and such that
\begin{equation}
f(Z;a),g(Z;b)~\mid~D=1=E~~~~~\dot\sim~~~~~X,Y.
\end{equation}
\end{mytask}
\noindent The symbol `$|$' stands for `conditional on', and `$\dot\sim$'  means 
`is approximately distributed as'.
The quality of the approximation needs to be
quantified; in our case, the supremum over $a$ and $b$ of the
variation distance between the two probability laws could be convenient
(a low score means high quality).
Moreover, one would like to have a uniformly large chance of acceptance.
Thus a further interesting score (high score means high quality) is
$\inf_{a,b}\Pr\{D=1=E\}$.

\section{The Solutions}
By \citet{bell} there is no way to succeed in Task 1 for Case 1. Moreover, there is no
way to succeed in Task 1 for Case 2 either, for certain suitably chosen two-point
sets of values for $a$ and $b$. 

Consider now Task 2, and suppose first of all that there are only two possible different
values of $a$ and $b$ each (Case 2). Let the random variable $Z$ consist of
independent coin tosses coding guesses for $a$ and $b$, and a realization of the pair
$X,Y$ drawn from the \emph{guessed} joint distribution. 
The transformations $\delta$ and $\epsilon$ check if each guess is correct. The
transformations $f$ and $g$ simply deliver the already generated $X$, $Y$. One obtains perfect
accuracy with success probability $1/4$. It is known that a much higher success probability is
achievable at the expense of more complicated transformations.

Now consider Task 2 for Case 1. So there is a continuum of possible values of $a$
and $b$. Note that the joint law of $X,Y$ depends on the parameters $a$, $b$ continuously,
and the parameters vary in compact sets. 
So one can partition each of their ranges into a finite number of cells
in  such a way that the joint law of 
$X,Y$ does not change much while each parameter varies within one cell of their respective
partitions. Moreover, one can get less and less variation at the expense of more and more cells.
Pick one representative parameter value in each cell.
 
Now, fix one of these pairs of partitions, and just play the obvious generalization of our
guessing game, using the representative parameter values for the guessed cells. 
If each partition has $k$ cells and the guesses are uniform and independent,
our success probability is $1/k^2$, uniformly in $a$ and $b$. We can achieve arbitrarily
high accuracy, uniformly in $a$ and $b$, at the cost of arbitrarily low success probability.
 
However, \citet{gisin2} show we can do much better in the case of the singlet correlations:
 \begin{mytheorem}\textbf{Perfect conditional simulation of the singlet correlations.}
For Case 1 and Task 2, there exists a perfect simulation with success probability
uniformly equal to $1/2$. 
\end{mytheorem}
 \noindent See \citet{gisin2} for the very pretty details. 
Can we do better still? What is the maximum uniformly achievable
success probability?
 
The joint laws coming from quantum mechanics always satisfy \emph{no action at a
distance} (``no Bell telephone''), i.e., the marginal of $X$ does not depend on 
$b$ nor that of $Y$ on $a$.
This should obviously be favourable to finding solutions to our tasks. Does it indeed play
a role in making these simulations spectacularly more easy for quantum mechanics, 
than in general?
Does ``no action at a distance'' ensure that we can find a perfect solution to
Task 2 with success probability uniformly bounded away from $0$? 
Am I indeed correct in thinking that one find probability distributions $p$
\emph{with} action at a distance, 
depending smoothly on parameters $a$, $b$, for which one can only achieve
perfection in the limit of zero success probability?

It would be interesting to study these problems in a wider context: arbitrary
biparameterized joint laws $p$; extend from pairs to triples; \dots

\section{Variant 1: Coincidences}
 
Instead of demanding that $\delta$ and $\epsilon$ in Task 2 are binary, one might allow
them to take on arbitrary real values, and correspondingly allow a more rich 
acceptance rule. Suggestively changing the notation to suggest times, define
now $S=\delta(Z;a)$ and $T=\epsilon(Z;b)$. Instead of conditioning on the separate events
$D=1$ and $E=1$ condition on the event $|S-T|<c$ where $c$ is some constant.
Obviously the new variant contains the original, so Variant Task 2 is at least as easy
as the original. 
Accardi, Imafuku and Regoli  (\citeyear{accardietal}) suggest that they
tackle this variant task 
claiming that it has nothing to do with
detector efficiency, but on the contrary is intrinsic to quantum optics, that one
must post-select on coincidences in arrival times of entangled photons.
By Heisenberg uncertainty, photons will always have a chance to arrive 
(or to be measured) at different times.
In those cases their joint state is not the singlet state. Therefore, 
if we were to collect data on all pairs (supposing $100\%$ detector efficiency)
we would not recover the singlet correlations. 

In actual fact the mathematical model of Accardi {\it et al.}  (2002, 2003)
applies to the original task, not the variant. Still, in many experiments this
kind of coincidence post-selection is done. Its effects (in terms of
the loophole issue) has never yet been analysed. The common 
consensus is that it is no worse than the usual detection loophole.
I convert this consensus into a conjecture:

 \begin{myconjecture}\textbf{No improvement from coincidences.}
There is no gain from Variant Task 2 over the original.
\end{myconjecture}
 \noindent Amazingly, this conjecture turns out to be \emph{false}. In quantitative terms
 the ``\emph{coincidence loophole}'' is about twice as serious
 as the detection loophole; see Larsson and Gill (2003).
 Fortunately, modern experimenters are moving (as Bell, 1981, stipulated)
 toward pulsed experiments
 and/or to event-ready detectors. In such an experment the detection time windows 
 are fixed in advance, not determined by the arrival times of the photons themselves.
 
 There seems to be a connection with the work of Massar, Bacon, Cerf and Cleve 
(\citeyear{massaretal}) on 
classical simulation of quantum entanglement using classical communication.
After all, checking  the inequality $|S-T|<c$ is a task which requires communication
between the two observers.

\section{Variant 2: Demanding More}

Instead of making Task 2 easier, as in the previous section, we can try to
make it harder by demanding further attractive properties of the simulated joint
probability distribution of $D,X,E,Y$.  For instance,
\citet{gisin2}  show how one can achieve
nice symmetry and stochastic independence properties at the cost of an
only slightly smaller success probability $4/9=(2/3)^2$. In fact, this solution 
has even more nice properties, as follows.

One might  like the simulated $X$ to behave well, when $D=1$,
whether or not $E=1$, and similarly for $Y$. 

Suppose we start with a joint law of $X,Y$ depending on $a$, $b$ as before.
Let $\eta$ be a fixed probability. Modify Task 2 as follows: we require not
only that given $D=1=E$, the simulated $X,Y$ have the prespecified joint
distribution, but also that conditional on $D=1$ and $E=0$, the simulated $X$ has the
prespecified marginal distribution, and also that, 
conditional on $D=0$ and $E=1$, the simulated $Y$
has the prespecified marginal distribution, and also that 
$D$ and $E$ are independent
Bernoulli$(\eta)$. Another way to describe this is by saying that under the
simulated joint probability distribution of $X,D,Y,E$, we have
statistical independence between $D$, $E$, and $(X,Y)$, with $(X,Y)$
distributed according to our target distribution and
$D$ and $E$ Bernoulli$(\eta)$, except that we don't care
about $X$ on $\{D=0\}$ nor about $Y$ on $\{E=0\}$

\citet{gisin2} show that this Variant Task 2 can be achieved for our main example 
Case 1, with $\eta=2/3$. It is known from considerations of the \citet{clauserhorne}
inequality that it cannot be done with $\eta>2/(1+\sqrt 2)\approx 0.828$. 
It seems that the precise boundary is unknown.

In fact, for some practical applications, achieving this task is more than necessary.
A slightly more modest task is to simulate the joint probability distribution just
described, conditionally on the complement of the event  $\{D=0=E\}$,
i.e. conditional on $D=1$ or $E=1$. This means to say that we also
don't care what is the simulated probability of $\{D=0=E\}$.  
\citet{gisin2} show that this can be
achieved with a variant of the same model, and with success probability $100\%$ 
(i.e., the simulation never generates an event $\{D=0=E\}$), and $\eta=2/3$.

\section*{Acknowledgments}
I am grateful for the warm hospitality and support  of the Quantum Probability 
group at the department of mathematics of the University of Greifswald, 
Germany, during my sabbatical there, Spring 2003. 
My research there was supported by European Commission grant 
HPRN-CT-2002-00279, RTN QP-Applications.
This research has also been supported by project RESQ (IST-2001-37559) 
of the IST-FET programme of the European Commission.

\end{document}